\documentclass[preprint,prc,showpacs,preprintnumbers,amsmath,amssymb,floatfix]{revtex4-2}
\usepackage{amsmath}
\usepackage{graphicx}
\usepackage{dcolumn}
\usepackage{bm}
\usepackage{ulem} 
\usepackage[usenames]{color}
\usepackage{epstopdf}
\usepackage{epsfig}
\usepackage{float}
\usepackage{subfigure}   
\usepackage{booktabs}
\usepackage{siunitx}
\usepackage{multirow}
\usepackage{physics}
\usepackage{xcolor} 
\usepackage{multirow}      
\usepackage[colorlinks,linkcolor=blue,anchorcolor=teal,citecolor=red]{hyperref}

\newcommand{\beq}{\begin{equation}}
	\newcommand{\eeq}{\end{equation}}
\newcommand{\beqa}{\begin{eqnarray}}
	\newcommand{\eeqa}{\end{eqnarray}}
\newcommand{\be}{\begin{eqnarray}}
	\newcommand{\ee}{\end{eqnarray}}

\begin{document}
	
	\title{Quantum information in neutron-proton scattering from the $M$ matrix}
	
	\author{Linjun Xie}
	\affiliation{School of Physics, Nankai University, Tianjin 300071, China}
	
	\author{Jinniu Hu}
	\email{hujinniu@nankai.edu.cn}
	\affiliation{
		School of Physics, Nankai University, Tianjin 300071, China\\
		and Shenzhen Research Institute of Nankai University, Shenzhen 518083, China
	}
	
	\author{Ying Zhang}
	\email{yzhangjcnp@tju.edu.cn}
	\affiliation{Department of Physics, School of Science, Tianjin University, Tianjin 300072, China}
	
	\author{Hong Shen}
	\affiliation{School of Physics, Nankai University, Tianjin 300071,  China}
	
	\date{\today}
	
	\newpage

	\begin{abstract}
		We study quantum-information aspects of neutron--proton scattering in the spin-space $M$-matrix framework. Four representative classes of input states are considered, namely diagonal mixed states, separable pure states, general two-qubit pure states, and a special Schmidt-like entangled subclass. For each class, ensemble-averaged output mutual information, reduced-state linear entropy, negativity, and geometric quantum discord are calculated in the relative momentum-- scattering angle plane. The results show that the outgoing spin correlations are governed jointly by scattering kinematics and by the structure of the incoming quantum ensemble. Input states with stronger intrinsic coherence or entanglement give larger maxima and higher minima in the mutual information, negativity, and geometric quantum discord. The enhanced regions of the mutual information and geometric discord depend on the input states, while the negativity maximum remains concentrated in the high-momentum backward-scattering region. These results extend earlier studies based on product-state entanglement power and provide an ensemble-based description of how spin correlations in neutron--proton scattering arise from the interplay between input-state structure and scattering dynamics.
	\end{abstract}
	
	\maketitle
	
	\section{Introduction}
	
	Quantum information has become a useful language for investigating correlations in quantum many-body systems. General concepts of many-body entanglement and quantum correlations are now well established from the quantum-information perspective \cite{Amico2008RMP,Horodecki2009RMP}. In nuclear physics, information-theoretic ideas appeared early in studies of reaction dynamics, for example in the use of information entropy to diagnose the nuclear liquid-gas phase transition \cite{Ma1999PRL}. More recently, entanglement-based measures have been used to discuss symmetries in chiral nucleon-nucleon potentials, to reorganize nuclear many-body wave functions, to optimize basis choices, and to develop truncation strategies for classical and quantum simulations.
	
	These developments now cover several complementary directions. In nuclear structure, entanglement entropy has been used to investigate cluster delocalization, self-consistent rearrangement, seniority, shell structure, proton-neutron correlations, and ab initio nuclear systems \cite{KanadaEnyo2015PRC,Robin2021PRC,Kruppa2022PRC,Tichai2023PLB,Johnson2023JPhysG,Gu2023PRC}. Measures of complexity and mode entanglement have also been introduced for many-fermion systems and for correlations associated with single-particle occupation probabilities and short-range correlations \cite{Bulgac2023PRC,Bulgac2023PRCLetter,Kou2024PLB}. In parallel, quantum magic, or nonstabilizerness, has been used to complement entanglement when assessing the quantum resources needed for nuclear simulations \cite{Brokemeier2025PRC}. These works show that entanglement alone does not fully characterize the computational complexity of nuclear wave functions.
	
	Quantum-information measures have also been extended to collective motion and reactions. Recent studies have examined entanglement and coherence in wobbling motion, entanglement in multinucleon transfer, quantum entanglement in fission, and entanglement between shape and rotation in low-lying nuclear states \cite{Chen2024PRC,Li2024PRC,Qiang2025PLB,Wang2025PRCLetter}. Related work further analyzed entanglement in two-quasiparticle triaxial rotor systems and proposed entanglement entropy as a signature of nuclear shape evolution \cite{Chen2025PRC,Wang2026PRC,Wan2026ovn}. Taken together, these studies indicate that quantum-information measures are becoming useful probes of nuclear structure, reactions, and many-body complexity.
	
	Scattering processes provide a particularly clean arena for these ideas. From the quantum-information point of view, scattering maps an incoming spin state to an outgoing spin state whose correlations are shaped by the interaction and by the symmetries of the scattering operator. Early work on dynamical entanglement in particle scattering emphasized that the amount of generated entanglement is constrained by boundary conditions and by the decomposition of the incoming state into different irreducible sectors \cite{Harshman2005IJMPA,Harshman2007IJQI}. Related studies in relativistic and high-energy scattering have examined the Lorentz-frame dependence of spin entanglement, tree-level entanglement generation in QED processes, and the broader opportunities for quantum-information measurements at colliders \cite{Fan2018PRD,Fedida2023PRD,Afik2025EPJP}. These works reinforce the view that scattering amplitudes can be used as microscopic maps for producing and studying quantum correlations.
	
	In low-energy strong interactions, the connection between entanglement and symmetry has become especially important. The entanglement power of the baryon-baryon \(S\) matrix was related to emergent spin-flavor symmetries, and entanglement suppression in the low-energy sector was shown to correlate with symmetry enhancement, including Wigner's SU(4) symmetry in the two-flavor case \cite{Beane2019PRL,LowMehen2021PRD,low2026quantum}. This idea was later extended to low-energy QCD with spin-\(1/2\) baryon octets, where minimal entanglement was connected with increasingly large accidental symmetries in the effective field theory \cite{LiuLowMehen2023PRC}. Similar analyses of hyperon-nucleon scattering and \(\Omega\Omega\) scattering indicate that entanglement suppression can also constrain strange-baryon interactions and spin-\(3/2\) channels \cite{LiuLow2024PLB,Hu2025Spin32Baryon,Sone2026OmegaOmega}.
	
	The scope of scattering-based quantum-information studies has continued to broaden beyond low-energy nuclear scattering. For example, polarized electron-proton scattering at the Electron-Ion Collider has been proposed as a setting in which transverse spin preparation can generate entanglement and magic, with the deep-inelastic signal governed by transversity parton distributions \cite{Cheng2025EIC}. These developments reinforce the central role of spin preparation and kinematic selection in shaping the quantum correlations generated by scattering amplitudes.
	
	Neutron--proton (\(np\)) scattering is the most direct two-body nuclear process in which these ideas can be explored quantitatively. As a fundamental microscopic input of low-energy nuclear physics, nucleon-nucleon scattering constrains realistic interactions and underlies a wide range of nuclear structure and reaction calculations. From the quantum-information perspective, it provides the simplest setting in which spin correlations generated by the strong interaction can be isolated in a controlled way. Studies of few-nucleon and \(np\) scattering have clarified how entanglement depends on momentum, scattering angle, and initial spin configuration \cite{Bai2022PRC,Bai2023PLB}. Partially polarized mixed inputs and experimentally accessible protocols for determining two-nucleon spin entanglement have also been discussed \cite{Bai2023PRCmixed,Bai2024PRC}. Recent studies further related spin entanglement to accidental Wigner SU(4) and Serber symmetries in chiral effective field theory and showed that entanglement can probe the operator and partial-wave content of nuclear interactions \cite{Cavallin2026PRC}.
	
	Very recent studies have shown that fixed-angle nucleon-nucleon scattering amplitudes can possess particularly transparent Bell-state structures. In unpolarized \(pp\) elastic scattering, Shen et al. found a near-pure Bell-triplet state around $E_{\rm lab}=151~{\rm MeV}$ and $\theta_{\rm cm}=90^\circ$, where the spin amplitude is approximately a Bell-state transition
	operator \cite{Shen2025BellTriplet}. Wita{\l}a confirmed this mechanism and further showed that unpolarized proton-deuteron breakup can generate Bell-like proton pairs in quasifree and final-state-interaction configurations \cite{Witala2026Teleportation}. These results emphasize the role of kinematic selection and the internal spin structure of the scattering amplitude. 
	
	These developments further motivate a closer examination of the role of the input
	spin state. In the deuteron ground state, neutron-proton spin correlations have been studied by tracing out the orbital and isospin degrees of freedom, and strong spin entanglement was found for the $M=0$ projection and for coherent superpositions of angular-momentum projections \cite{Singh2025DeuteronSpin}. In elastic polarized \(np\) scattering, final polarization states have been reconstructed from the outgoing polarizations and spin-correlation coefficients, and most final states were found to be statistical mixtures under standard polarization preparation \cite{Witala2025FinalPolarization}. This latter result highlights the importance of distinguishing experimentally prepared polarization ensembles from more general input quantum states used to probe the spin-space scattering map.
	
	Despite this progress, most existing studies of spin entanglement in nucleon scattering have focused on restricted input states or on a limited set of entanglement-related quantities. The standard entanglement power, for example, is defined by averaging over initially unentangled product states \cite{Beane2019PRL,Bai2023PLB,Cavallin2026PRC}. It is well suited for characterizing the entangling capacity of a scattering operator, but it does not show how diagonal mixed states, separable pure states, generic two-qubit pure states, or controlled entangled subclasses respond to the same spin-space scattering amplitude. A broader comparison is also useful because mutual information, reduced-state linear entropy, negativity, and geometric discord characterize different aspects of quantum correlations \cite{Horodecki2009RMP,Brokemeier2025PRC,RobinSavage2025PRC,Cheng2025EIC,Liu2026QST}.
	
	The present work addresses this problem within a unified \(M\)-matrix framework for \(np\) scattering. At fixed relative momentum and scattering angle, we construct the normalized conditional output spin state and calculate the ensemble-averaged output mutual information, reduced-state linear entropy, negativity, and geometric quantum discord for several representative input-state classes. This comparison separates the effects of scattering kinematics from those of input-state structure at the level of averaged responses, and extends earlier analyses based mainly on product-state entanglement power, selected spin configurations, or experimentally prepared polarization ensembles \cite{Beane2019PRL,Bai2022PRC,Bai2023PLB,Bai2023PRCmixed,Cavallin2026PRC,Witala2025FinalPolarization}. It provides a systematic picture of how the incoming spin ensemble shapes the quantum-information content of the outgoing spin state.
	
	This paper is organized as follows. In Sec.~II, we introduce the spin-space scattering framework and define the input-state classes and quantum-information measures used in this work. In Sec.~III, we present the numerical results and discuss the ensemble-averaged correlation patterns in the momentum-angle plane. Finally, Sec.~IV summarizes the main conclusions and outlines possible extensions.
	
	\section{Theoretical formalism}
	
	\subsection{Input-state classes}
	
	For \(np\)  scattering, the input spin state is defined in the four-dimensional basis
	\[
	\{ \ket{n_\uparrow p_\uparrow},\ \ket{n_\uparrow p_\downarrow},\ \ket{n_\downarrow p_\uparrow},\ \ket{n_\downarrow p_\downarrow} \},
	\]
	where \(n_\uparrow\), \(n_\downarrow\), \(p_\uparrow\), and \(p_\downarrow\) denote neutron and proton spin-up and spin-down states, respectively. The input spin state is represented by a density matrix \(\rho_{\rm in}\), with the following four representative classes considered in this work.
	
	A diagonal mixed state is first introduced,
	\begin{equation}
		\rho_{\text{in}}
		= \operatorname{diag}\{ p_1, p_2, p_3, p_4 \},
		\qquad
		p_1+p_2+p_3+p_4=1,
		\qquad
		p_i\geq 0,
		\label{eq:input_mixed}
	\end{equation}
	which is a separable mixed state representing a classical probabilistic mixture of the four product-basis states. Depending on the choice of \(\{p_i\}\), this state may still contain classical correlations, but it contains no quantum coherence in the chosen product basis.
	
	We next include the separable pure state. In Bloch-sphere form, a general separable two-qubit pure state may be written as \cite{Bai2023PLB}
	\begin{equation}
		\psi_{\text{in}}
		=
		\begin{bmatrix}
			\cos(\theta_1/2) \\
			e^{i\varphi_1}\sin(\theta_1/2)
		\end{bmatrix}
		\otimes
		\begin{bmatrix}
			\cos(\theta_2/2) \\
			e^{i\varphi_2}\sin(\theta_2/2)
		\end{bmatrix}
		=
		\begin{bmatrix}
			\cos(\theta_1/2) \cos(\theta_2/2) \\
			e^{i\varphi_2} \cos(\theta_1/2) \sin(\theta_2/2) \\
			e^{i\varphi_1} \sin(\theta_1/2) \cos(\theta_2/2) \\
			e^{i(\varphi_1 + \varphi_2)} \sin(\theta_1/2) \sin(\theta_2/2)
		\end{bmatrix},
		\label{eq:input_separable}
	\end{equation}
	where \(\theta_{1},\theta_{2}\in[0,\pi]\) and \(\varphi_{1},\varphi_{2}\in[0,2\pi)\). This state contains initial spin coherence in each subsystem, but it contains no initial entanglement between the neutron and proton spins.
	
	To represent arbitrary pure input states, we use the following six-parameter Schmidt-type form \cite{NielsenChuang2010,BengtssonZyczkowski},
	\begin{equation}
		\psi_{\text{in}}
		=
		\cos\frac{\chi}{2}
		|\Psi(\theta_1,\varphi_1)\rangle
		\otimes
		|\Psi(\theta_2,\varphi_2)\rangle
		+
		e^{i\delta}\sin\frac{\chi}{2}
		|\Psi^\perp(\theta_1,\varphi_1)\rangle
		\otimes
		|\Psi^\perp(\theta_2,\varphi_2)\rangle,
		\label{eq:input_general_pure}
	\end{equation}
	where
	\begin{eqnarray}
		|\Psi(\theta,\varphi)\rangle
		&=&
		\cos(\theta/2)|0\rangle
		+
		e^{i\varphi}\sin(\theta/2)|1\rangle,\\\nonumber
		|\Psi^\perp(\theta,\varphi)\rangle
		&=&
		-e^{-i\varphi}\sin(\theta/2)|0\rangle
		+
		\cos(\theta/2)|1\rangle.
		\label{eq:single_qubit_orthogonal}
	\end{eqnarray}
	Here \(\chi\in[0,\pi]\) controls the Schmidt weights and \(\delta\in[0,2\pi)\) is the relative phase between the two Schmidt branches. Together with the two local Bloch-sphere directions, these parameters provide a convenient representation of arbitrary two-qubit pure states for ensemble averaging.
	
	Finally, in addition to the full pure-state class above, we introduce a special entangled subclass for comparison,
	\begin{equation}
		\psi_{\text{in}}
		=
		\begin{bmatrix}
			\cos\alpha\cos\beta \\
			e^{i\vartheta}\cos\alpha\sin\beta \\
			e^{i\phi}\sin\alpha\sin\beta \\
			-e^{i(\vartheta+\phi)}\sin\alpha\cos\beta
		\end{bmatrix},
		\qquad
		\alpha,\beta\in\left[0,\frac{\pi}{2}\right],
		\qquad
		\vartheta,\phi\in[0,2\pi).
		\label{eq:input_schmidt_like}
	\end{equation}
	This subclass is not intended to parametrize the full two-qubit pure-state space. It serves as a controlled entangled subclass used to contrast the responses of intrinsically entangled inputs with those of the other input-state classes. The motivation for this choice is that its intrinsic entanglement is analytically transparent. For this subclass, one finds \(ad-bc=-e^{i(\vartheta+\phi)}\cos\alpha\sin\alpha\), and the pure-state concurrence is therefore
	\begin{equation}
		C=2|ad-bc|=\sin 2\alpha.
		\label{eq:schmidt_like_concurrence}
	\end{equation}
	The parameter \(\alpha\) continuously tunes the state from a product state to a maximally entangled state, while \(\beta\) and the phases \(\vartheta\) and \(\phi\) vary the spin composition and relative phases inside the same controlled subclass. This makes Eq.~\eqref{eq:input_schmidt_like} a useful benchmark between the separable pure-state class and the full pure-state class. It is narrower than a Haar-random two-qubit pure-state ensemble, but it is richer and less basis-specific than a single Bell-state input.
	
	\subsection{Output state}
	The structure of the spin-space scattering \(M\) matrix from nucleon-nucleon scattering is constrained by Lorentz symmetry, rotational invariance, parity, time-reversal invariance, and the unitarity of the full scattering matrix. Under these constraints, it can be written in terms of six complex amplitudes that depend on the energy and scattering angle.  The six quantities \(M_1,\ldots,M_6\) are the independent complex spin amplitudes of elastic \(np\) scattering at the chosen momentum and scattering angle. They should be understood as transition amplitudes between the initial and final two-nucleon spin states in the basis specified above.  For example, \(M_1\) corresponds to the spin-nonflip matrix element in the parallel-spin sector \cite{Wolfenstein1954,Bystricky1978, Nan2024NNPotentials},
	\begin{equation}
		M_1
		=
		\langle n_\uparrow p_\uparrow |M| n_\uparrow p_\uparrow\rangle
		=
		\langle n_\downarrow p_\downarrow |M| n_\downarrow p_\downarrow\rangle .
	\end{equation}
	In this basis, \(M_1\) and \(M_3\) describe spin-nonflip transitions in the parallel-spin and antiparallel-spin sectors, respectively. 
	The amplitudes \(M_2\) and \(M_4\) connect states in which both spin projections are changed. 
	Thus they represent double-spin-flip or spin-exchange structures. 
	The amplitudes \(M_5\) and \(M_6\) connect states that differ by one spin projection and therefore encode the two independent single-spin-flip structures allowed for neutron-proton scattering.  {In the exchange-symmetric or isospin-invariant limit adopted in Ref, \cite{Bystricky1978}, one has \(M_5=-M_6\) .  However, the two single-helicity-flip amplitudes are in general independent in high-precision nuclear potentials due to the isospin-symmetry breaking as discussed in Ref.  \cite{bystricky1984tests}, which is the convention adopted in NN-online \cite{NNOnline}.}
	
	All spin observables are built from these amplitudes. In the spin basis used here, we write
	\begin{equation}
		M = \begin{pmatrix}
			M_1 & M_5 & M_6 & M_2 \\
			M_6 & M_3 & M_4 & M_6 \\
			M_5 & M_4 & M_3 & M_5 \\
			M_2 & M_5 & M_6 & M_1
		\end{pmatrix}.
		\label{eq:M_matrix}
	\end{equation}

	For a pure input state \(\psi_{\text{in}}\), we define the normalized conditional output state at fixed kinematics following the standard nonforward-scattering construction \cite{Bai2023PLB,Cavallin2026PRC},
	\begin{equation}
		|\psi_{\text{out}}\rangle
		=
		\frac{M |\psi_{\text{in}}\rangle}
		{\sqrt{\langle \psi_{\text{in}} | M^\dagger M | \psi_{\text{in}} \rangle}} .
		\label{eq:pure_output_state}
	\end{equation}
	The corresponding density matrix \(\rho_{\text{out}}=\ket{\psi_{\text{out}}}\bra{\psi_{\text{out}}}\) is
	\begin{equation}
		\rho_{\rm out}
		=
		\sum_{i,j=1}^{4}
		\rho_{ij}
		|i\rangle\langle j|,
		\qquad
		\{|i\rangle\}
		=
		\{|n_\uparrow p_\uparrow\rangle,
		|n_\uparrow p_\downarrow\rangle,
		|n_\downarrow p_\uparrow\rangle,
		|n_\downarrow p_\downarrow\rangle\}.
	\end{equation}
	
	For mixed-state inputs as shown in Eq. \eqref{eq:input_mixed}, the corresponding conditional output density matrix is obtained directly from \(\rho_{\text{in}}\) as
	\begin{equation}
		\rho_{\text{out}}
		=
		\frac{M\rho_{\text{in}}M^\dagger}
		{\operatorname{Tr}[M\rho_{\text{in}}M^\dagger]}.
		\label{eq:mixed_output_state}
	\end{equation}
	In both the pure- and mixed-state cases, the normalization ensures that we work with a post-selected conditional outgoing spin state at fixed momentum and scattering angle. This map should not be interpreted as a complete trace-preserving quantum channel, because the kinematic selection has already been imposed.
	
	\subsection{Quantum-information observables}
	
	Quantum mutual information measures the total amount of correlations, including both classical and quantum contributions, contained in a bipartite state \cite{NielsenChuang2010,Groisman2005}. For the conditional output state, it is defined as
	\begin{equation}
		\mathcal{I}_{\text{out}}(n:p)=S(\rho_n)+S(\rho_p)-S(\rho),
		\label{eq:mutual_information}
	\end{equation}
	where \(S(\rho)=-\mathrm{Tr}(\rho\log\rho)\) is the von Neumann entropy, and \(\rho_n=\mathrm{Tr}_p(\rho)\) and \(\rho_p=\mathrm{Tr}_n(\rho)\) are the reduced density matrices of the neutron and proton spin subsystems. In terms of the matrix elements of \(\rho\), they are
	\begin{equation}
		\rho_{\text{n}}
		=
		\begin{pmatrix}
			\rho_{11}+\rho_{22} & \rho_{13}+\rho_{24} \\
			\rho_{31}+\rho_{42} & \rho_{33}+\rho_{44}
		\end{pmatrix},
		\qquad
		\rho_{\text{p}}
		=
		\begin{pmatrix}
			\rho_{11}+\rho_{33} & \rho_{12}+\rho_{34} \\
			\rho_{21}+\rho_{43} & \rho_{22}+\rho_{44}
		\end{pmatrix}.
		\label{eq:reduced_density_matrices}
	\end{equation}
	
	For a given density matrix \(\rho\), the reduced-state linear entropy is defined by \cite{Zanardi2000,Bai2023PLB}
	\begin{equation}
		S_L(\rho_A)=1-\mathrm{Tr}(\rho_A^2).
		\label{eq:linear_entropy}
	\end{equation}
	It measures the mixedness of the reduced state. When the full bipartite state \(\rho_{AB}\) is pure, \(S_L(\rho_A)\) is a standard proxy for bipartite entanglement. When the full output state is mixed, however, \(S_L\) should be interpreted more cautiously as a reduced-state mixedness measure rather than as an entanglement measure in its own right.
	
	To quantify entanglement for both pure and mixed output states, we use the Peres--Horodecki positive-partial-transpose criterion \cite{Peres1996,Horodecki1996}. For a bipartite system with orthonormal basis \(\{|k_n m_p\rangle=|k_n\rangle\otimes |m_p\rangle\}\), a general density matrix \(\rho_{np}\) can be parameterized as
	\begin{equation}
		\begin{aligned}
			& \rho_{np} 	= \sum_{kmln}(\rho_{np})_{k_n m_p;l_n n_p}
			|k_n m_p\rangle\langle l_n n_p|, \\
			& (\rho_{np})_{k_n m_p;l_n n_p} = \langle k_n m_p|\rho_{AB}|l_n n_p\rangle .
		\end{aligned}
		\label{eq:rho_np_expansion}
	\end{equation}
	Its partial transpose \(\rho_{np}^{T_n}\) is defined as
	\begin{align}
		\rho_{np}^{T_n}
		=
		\sum_{kmln}(\rho_{np})_{l_n m_p;k_n n_p}
		|k_n m_p\rangle\langle l_n n_p| .
		\label{eq:partial_transpose}
	\end{align}
	We then use the negativity \cite{Zyczkowski1998,VidalWerner2002}
	\begin{equation}
		\mathcal{N}(\rho)
		=
		-\sum_{\lambda_i<0}\lambda_i,
		\label{eq:negativity}
	\end{equation}
	to quantify entanglement, where \(\lambda_i\) are the negative eigenvalues of \(\rho_{np}^{T_n}\).
	
	While mutual information measures the total correlation, quantum discord is designed to capture nonclassical correlations that may remain even when the state is separable. The classical contribution is extracted through local projective measurements on one subsystem and is written as \(\mathcal{J}(n:p)=\max_{\{\Pi_i\}}\mathcal{I}(n:p\mid\{\Pi_i\})\). When subsystem \(n\) is measured, the quantum discord \cite{OllivierZurek2001,HendersonVedral2001} is
	\begin{equation}
		\mathcal{D}(p:n)=\mathcal{I}(n:p)-\mathcal{J}(n:p).
		\label{eq:discord}
	\end{equation}
	When \(\mathcal{D}(p:n)>0\), the state contains quantum correlations beyond a purely classical description. Such correlations may exist even when the negativity is zero.
	
	The original definition of quantum discord involves a nontrivial optimization over local measurements and is often difficult to evaluate analytically. We therefore employ geometric quantum discord \cite{Dakic2010,LuoFu2010}, which measures the distance between a given quantum state \(\rho_{np}\) and the nearest classically correlated state \(\chi\),
	\begin{equation}
		\mathcal{D}_{G}(\rho_{np})
		=
		\frac{1}{m-1}\min_{\chi\in\Omega_c}
		\|\rho_{np}-\chi\|^2,
		\label{eq:geometric_discord_definition}
	\end{equation}
	where \(\|X\|^2=\operatorname{Tr}(X^\dagger X)\) is the square of the Hilbert-Schmidt norm.
	
	Any two-qubit density matrix can be parameterized using the Bloch representation as
	\begin{equation}
		\rho_{np}
		=
		\frac{1}{4}
		\sum_{\mu=0}^{3}\sum_{\nu=0}^{3}
		\mathcal{T}_{\mu\nu}\,\sigma_\mu\otimes\sigma_\nu,
		\label{eq:bloch_representation}
	\end{equation}
	where \(\sigma_\mu\) are the Pauli-matrix basis operators and \(\mathcal{T}_{\mu\nu}\) are real coefficients. When subsystem \(n\) is measured, the geometric quantum discord has the analytic form
	\begin{equation}
		\mathcal{D}_{G}(p:n)
		=
		\frac{1}{4}
		\left(
		\sum_{j=1}^{3}\sum_{\nu=0}^{3}\mathcal{T}_{j\nu}^{2}
		-
		\lambda_{\max}
		\right),
		\label{eq:geometric_discord}
	\end{equation}
	where \(\lambda_{\max}\) is the largest eigenvalue of the \(3\times3\) matrix \(L_n=a a^T+E E^T\). Here \(a=(\mathcal{T}_{10},\mathcal{T}_{20},\mathcal{T}_{30})^T\) contains local information, while \(E_{ij}=\mathcal{T}_{ij}\) with \(i,j=1,2,3\) characterizes the two-body correlation tensor.
	
	\subsection{Ensemble averages and sampling measures}
	
	After defining the observables, we denote any one of them by
	\begin{equation}
		Q_{\rm out}
		\in
		\left\{
		\mathcal{I}_{\rm out},\,
		S_L,\,
		\mathcal{N},\,
		\mathcal{D}_{G}
		\right\}.
		\label{eq:Q_out_set}
	\end{equation}
	For each input-state class \(\mathcal{F}\), the ensemble-averaged output quantity is defined as
	\begin{equation}
		\overline{Q}_{\mathcal{F}}(p,\theta)
		=
		\int_{\mathcal{F}}d\mu_{\mathcal{F}}(\xi)\,
		Q_{\rm out}(p,\theta;\xi),
		\label{eq:ensemble_average}
	\end{equation}
	where \(\xi\) denotes the parameters of the input state and \(d\mu_{\mathcal{F}}\) is the sampling measure associated with the corresponding input-state class. The word ensemble refers to the statistical ensemble generated by the input-state class and its sampling measure. It does not denote an average over scattering events.
	
	For the diagonal mixed-state class, the average is taken over the probability simplex,
	\begin{equation}
		d\mu_{\rm mix}
		=
		6\,
		\delta\left(1-\sum_{i=1}^{4}p_i\right)
		\prod_{i=1}^{4}dp_i,
		\qquad
		p_i\geq0 .
		\label{eq:measure_mix}
	\end{equation}
	This is the uniform Dirichlet measure for the four probabilities.
	
	For the separable pure-state class, the average is taken over the product of two single-qubit Fubini--Study measures,
	\begin{equation}
		d\mu_{\rm sep}
		=
		\frac{\sin\theta_1\,d\theta_1\,d\varphi_1}{4\pi}
		\frac{\sin\theta_2\,d\theta_2\,d\varphi_2}{4\pi},
		\label{eq:measure_sep}
	\end{equation}
	where \(0\leq\theta_i\leq\pi\) and \(0\leq\varphi_i<2\pi\).
	
	For the general two-qubit pure-state class, the average is taken with the Fubini--Study measure on the two-qubit pure-state space \cite{ZyczkowskiSommers2001} written in the variables of Eq.~\eqref{eq:input_general_pure},
	\begin{equation}
		d\mu_{\rm pure}
		=
		\frac{\sin\theta_1\,d\theta_1\,d\varphi_1}{4\pi}
		\frac{\sin\theta_2\,d\theta_2\,d\varphi_2}{4\pi}
		\frac{d\delta}{2\pi}
		\frac{3}{2}\sin\chi\cos^2\chi\,d\chi,
		\label{eq:measure_pure}
	\end{equation}
	where \(0\leq\chi\leq\pi\) and \(0\leq\delta<2\pi\). The last factor gives the induced distribution of the Schmidt weights.
	
	For the special Schmidt-like entangled subclass, we adopt a normalized weighted measure in its prescribed parameter space,
	\begin{equation}
		d\mu_{\rm Sch}
		= 	\frac{\sin\alpha}{2\pi}\frac{\sin\beta}{2\pi} d\alpha\,d\beta\,d\vartheta\,d\phi,
		\label{eq:measure_schmidt_like}
	\end{equation}
	with \(0\leq\alpha,\beta\leq\pi/2\) and \(0\leq\vartheta,\phi<2\pi\). This measure is not intended to represent the Haar measure over all entangled two-qubit pure states. It defines a controlled average within the restricted subclass of Eq.~\eqref{eq:input_schmidt_like}.

	\section{Numerical results and discussion}
	
	For each kinematic point $(p,\theta)$, where $p$ denotes the relative momentum corresponding to the laboratory kinetic energy and $\theta$ denotes the scattering angle, we evaluate the output quantity after averaging over the corresponding input-state class. All figures in this section therefore show ensemble-averaged responses rather than the behaviour of a single input state. The Monte Carlo estimator is applied independently at each point of the kinematic plane. Its sample size is set to $N_{\rm MC}=6\times 10^5$ for each point. Further increases of the sample number do not produce visible changes in the contour patterns or in the extracted extrema. The spin-space \(np\) \(M\)-matrix elements used in this work are taken from the PWA93 data from the Nijmegen partial-wave analysis \cite{Stoks1993NNPWA}. 
	The numerical amplitudes are obtained from the NN-OnLine amplitude database \cite{NNOnline}, which provides several equivalent amplitude representations, including Wolfenstein and helicity amplitudes. The extrema extracted from the contour plots are summarized in Table~\ref{table_1}.
	
	\begin{figure}[htbp]
		\centering
		\includegraphics[width=0.8\textwidth, keepaspectratio]{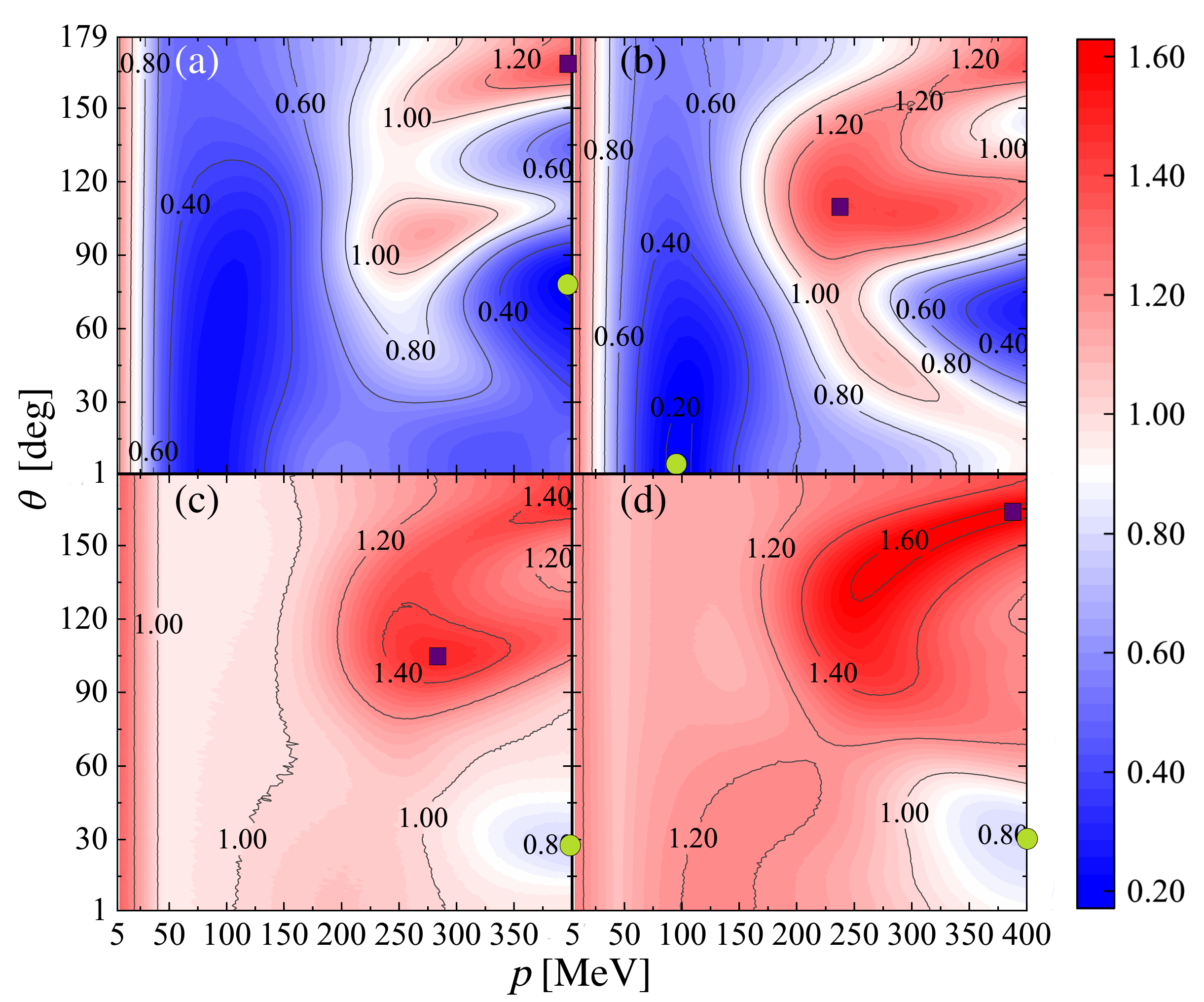}
		\caption{Momentum- and angle-dependent ensemble-averaged output mutual information $\overline{\mathcal I}(p,\theta)$. Panels (a)--(d) correspond to averages over diagonal mixed states, separable pure states, general two-qubit pure states, and the special Schmidt-like entangled subclass, respectively. The global maximum in each panel is marked by a purple square and the global minimum by a green circle.}
		\label{fig:QIP_Iout}
	\end{figure}
	
	The ensemble-averaged output mutual information $\overline{\mathcal I}(p,\theta)$ is presented in Fig.~\ref{fig:QIP_Iout}. The peak values differ systematically among the four input-state classes. The maximum increases from $1.370$ for the diagonal mixed-state class to ${1.404}$ for the separable class, $1.487$ for the general pure-state class, and $1.630$ for the Schmidt-like entangled subclass. Relative to the diagonal mixed-state class, these values correspond to enhancements of about $3\%$, $9\%$, and $19\%$. The minima are also strongly lifted for the coherent and entangled pure-state classes. The minima for the diagonal mixed and separable classes are $0.185$ and {$0.180$}, while those for the general and Schmidt-like pure-state classes are $0.786$ and $0.793$. Intrinsic coherence or entanglement in the input ensemble therefore enhances the strongest region and also raises the overall correlation baseline across the kinematic plane.
	
	The position of the largest response in the \((p,\theta)\) plane depends on the input-state class. For the diagonal mixed-state class, the global maximum lies in the high-momentum, large-angle region at $(399.6~\mathrm{MeV},167.0^\circ)$. For the separable pure-state class, the maximum shifts to {(237.5 MeV, 110.0$^\circ$)}.  {This location is close to the maximum of the product-state entanglement power reported for exact \(np\) scattering in Ref.~\cite{Bai2023PLB}}. This agreement is natural because the entanglement power is defined through an average over separable pure inputs. For the general pure-state class, the maximum moves to $(284.3~\mathrm{MeV},104.0^\circ)$. For the Schmidt-like entangled subclass, the maximum is shifted to the high-momentum, backward-angle region at $(390.1~\mathrm{MeV},164.0^\circ)$. The kinematic position of the strongest total-correlation response is therefore not universal. It changes when the input ensemble is enlarged from product states to coherent or intrinsically entangled classes.
	
	This behavior extends the picture obtained from earlier entanglement-power studies. In the $S$-wave limit, the entangling ability of the scattering operator is controlled by the relative phase between spin-singlet and spin-triplet sectors \cite{Harshman2005IJMPA,Harshman2007IJQI,Beane2019PRL}. Once the full angle-dependent $M$ matrix is used, the averaged total correlation is also sensitive to the input ensemble. The broad regions of elevated \(\overline{\mathcal I}\) indicate that incoming coherence can be converted into output correlation over an extended part of the kinematic plane. The Schmidt-like entangled subclass gives the most uniform high-correlation pattern among the four classes.
	
	\begin{figure}[htbp]
		\centering
		\includegraphics[width=0.8\textwidth, keepaspectratio]{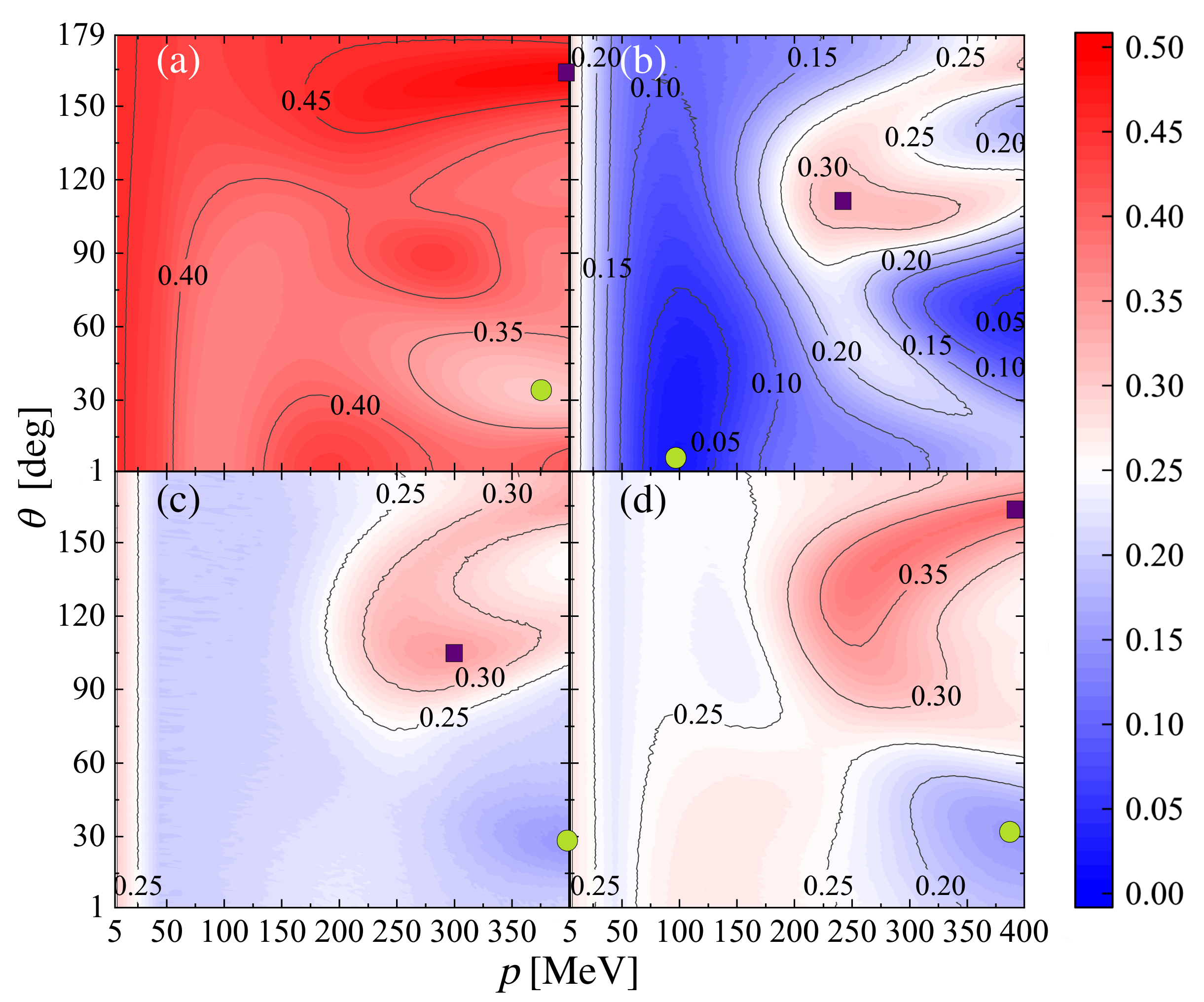}
		\caption{Momentum- and angle-dependent ensemble-averaged output linear entropy $\overline{S}_L(p,\theta)$.  Panels (a)--(d) correspond to averages over the same four input-state classes as in Fig.~\ref{fig:QIP_Iout}. Larger values indicate stronger reduced-state mixedness. For pure bipartite output states, they also track stronger entanglement.}
		\label{fig:QIP_SL}
	\end{figure}
	
	The ensemble-averaged output linear entropy $\overline{S}_L(p,\theta)$ is shown in Fig.~\ref{fig:QIP_SL}. Only the separable pure-state result in Fig.~\ref{fig:QIP_SL}(b) has the same interpretation as the standard product-state entanglement power used in Refs.~\cite{Beane2019PRL,Bai2023PLB}. The entanglement power is a property of a scattering operator obtained from product-state averaging. The linear entropy used here is a property of the reduced output state after averaging over a chosen input class. For pure bipartite output states, it is a useful proxy for entanglement. For mixed output states, \(S_L\) measures the mixedness of the reduced spin state and should not be interpreted directly as an entanglement measure.
	
	In Ref.~\cite{Bai2023PLB}, the entanglement power \(E(p,\theta)\) was defined as the average reduced-state linear entropy of the normalized output state generated from all initial product pure states.  This is precisely the quantity denoted here by \(\overline{S}_{L,\mathrm{sep}}(p,\theta)\), up to notation. The new element of the present work is that this product-state average is only one member of a broader comparison. 
	
	After reproducing the separable input result, we apply the same conditional \(M\)-matrix framework to diagonal mixed states, general two-qubit pure states, and a controlled Schmidt-like entangled subclass. 
	The comparison shows that the peak of \(\overline{S}_L\) increases from \(0.317\) for the separable class to \(0.386\) for the Schmidt-like class, while the corresponding enhanced regions shift toward backward angles at high momentum. 
	Thus the entanglement-power result of Ref.~\cite{Bai2023PLB} is recovered as the separable-pure-state sector of our analysis, and the present work extends it by quantifying how the averaged output correlations change when the input-state class is varied.
	
	The diagonal mixed-state class has both the largest maximum and the largest minimum among all four classes, with $\overline{S}_L^{\max}=0.492$ at $(397.2~\mathrm{MeV},164.0^\circ)$ and $\overline{S}_L^{\min}=0.304$ at $(375.4~\mathrm{MeV},33.0^\circ)$. For the diagonal mixed-state class, the large value of \(S_L\) mainly reflects the mixedness of the reduced one-particle spin state rather than stronger bipartite entanglement. The input state is already a statistical mixture of product spin states, and the conditional \(M\)-matrix map generally transforms these components into different output spin structures. After tracing over one nucleon, the reduced density matrix can therefore become close to a maximally mixed one-qubit state, for which
	\begin{equation}
		S_L
		=
		1-\mathrm{Tr}\left(\frac{I_2}{2}\right)^2
		=
		\frac{1}{2}.
	\end{equation}
	This explains why the diagonal mixed-state class has both a large maximum and a relatively large minimum of \(S_L\). 
	In contrast, pure-state inputs can still yield nearly pure reduced states in some kinematic regions, leading to much smaller minimum values of \(S_L\).
	
	The Schmidt-like entangled subclass gives the strongest averaged subsystem mixing within the pure-state sector. The separable and general pure-state classes peak at $(241.4~\mathrm{MeV},111.0^\circ)$ and $(300.3~\mathrm{MeV},105.0^\circ)$. These points lie in the same intermediate-to-large-angle region where the mutual information is enhanced. The Schmidt-like subclass instead peaks at $(397.2~\mathrm{MeV},164.0^\circ)$, close to the backward-scattering region observed in the mutual-information map. The linear entropy and mutual information therefore show related but not identical kinematic responses.
	
	The small-angle region suppresses $\overline{S}_L$ for the pure-state classes. For the separable class, the minimum appears at $(97.1~\mathrm{MeV},6.0^\circ)$. For the general and Schmidt-like subclasses, the minima occur at high momentum but remain at small angles, $(399.6~\mathrm{MeV},28.0^\circ)$ and $(387.7~\mathrm{MeV},32.0^\circ)$. This pattern suggests that forward scattering tends to preserve a comparatively simple reduced-state structure, while intermediate- and large-angle scattering produce stronger subsystem mixing.

	\begin{figure}[htbp]
		\centering
		\includegraphics[width=0.8\textwidth, keepaspectratio]{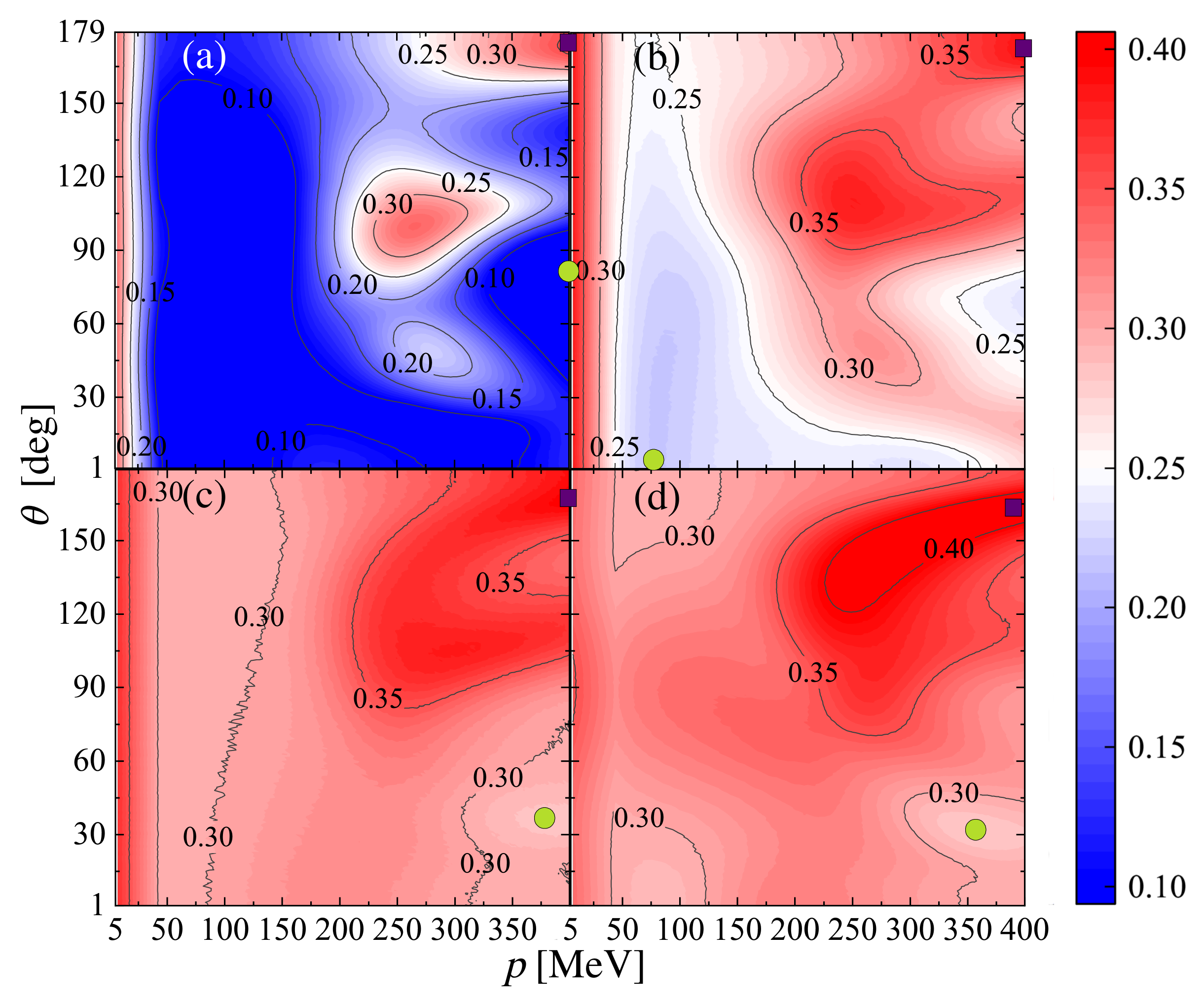}
		\caption{Momentum- and angle-dependent ensemble-averaged output negativity $\overline{\mathcal N}(p,\theta)$. The color scale is chosen to emphasize the high-value region. Panels (a)--(d) correspond to averages over the same four input-state classes as in Fig.~\ref{fig:QIP_Iout}.}
		\label{fig:QIP_Nega}
	\end{figure}
	
	The ensemble-averaged negativity $\overline{\mathcal N}(p,\theta)$ is given in Fig.~\ref{fig:QIP_Nega}. This quantity directly measures bipartite entanglement of the conditional output state. Compared with the mutual information and the linear entropy, the negativity has a more localized kinematic structure. The positions of the maxima are remarkably stable across the four input-state classes. For all four classes, the peak appears at large laboratory momenta and backward scattering angles. The most favourable region for producing or maintaining bipartite output entanglement is therefore concentrated in the high-momentum backward-scattering sector.
	
	This stability contrasts with the behaviour of the mutual information and geometric discord, whose positions change more strongly with the input class. It also complements the product-state entanglement-power analysis of Ref.~\cite{Bai2023PLB}, where the strongest averaged product-state entangling response occurs at intermediate momentum and large angle. The negativity result shows that the kinematic region that maximizes output entanglement for the averaged conditional state need not coincide with the region that maximizes the operator entangling power for separable inputs. This difference is expected because entanglement power and negativity answer different questions.
	
	Using the diagonal mixed-state class as the reference, \(\overline{\mathcal N}^{\max}\) increases by about \(9\%\), \(12\%\), and \(20\%\) for the separable, general pure, and Schmidt-like classes, respectively. The minima rise even more strongly. The value of $\overline{\mathcal N}^{\min}$ increases from $0.024$ for the diagonal mixed-state class to $0.217$ for the separable class and to about $0.284$ for the two classes containing stronger intrinsic coherence or entanglement. The pure-state classes therefore do not merely sharpen a local peak. They also suppress the low-entanglement region of the kinematic plane.
	
   The behavior of the negativity is different from that of the linear entropy because $\overline{\mathcal N}$ directly quantifies bipartite entanglement. For the diagonal mixed-state class, the input state contains only classical population weights in the product-spin basis and has no initial coherence or entanglement. Although the conditional \(M\)-matrix map can generate entanglement in some kinematic regions, the output entanglement is diluted by the statistical mixture of different input components. 
   As a result, the averaged negativity remains relatively small, especially in the low-entanglement regions.
   
   For pure-state inputs, the situation is different. The separable pure-state class contains local spin coherence, and the spin-flip and spin-exchange amplitudes in \(M\) can convert this coherence into bipartite output entanglement. The general pure-state class and the Schmidt-like subclass contain even stronger nonlocal coherence, since many of their input states are already entangled before scattering. In the Schmidt-like subclass, the intrinsic entanglement is controlled by $C=2|ad-bc|=\sin 2\alpha$.
   These initial nonlocal correlations are remixed and modulated by the scattering amplitudes rather than being removed by the \(M\)-matrix map. This explains why the pure and entangled input-state classes have larger averaged negativity and why their low-negativity regions are strongly suppressed.
	
	\begin{figure}[htbp]
		\centering
		\includegraphics[width=0.8\textwidth]{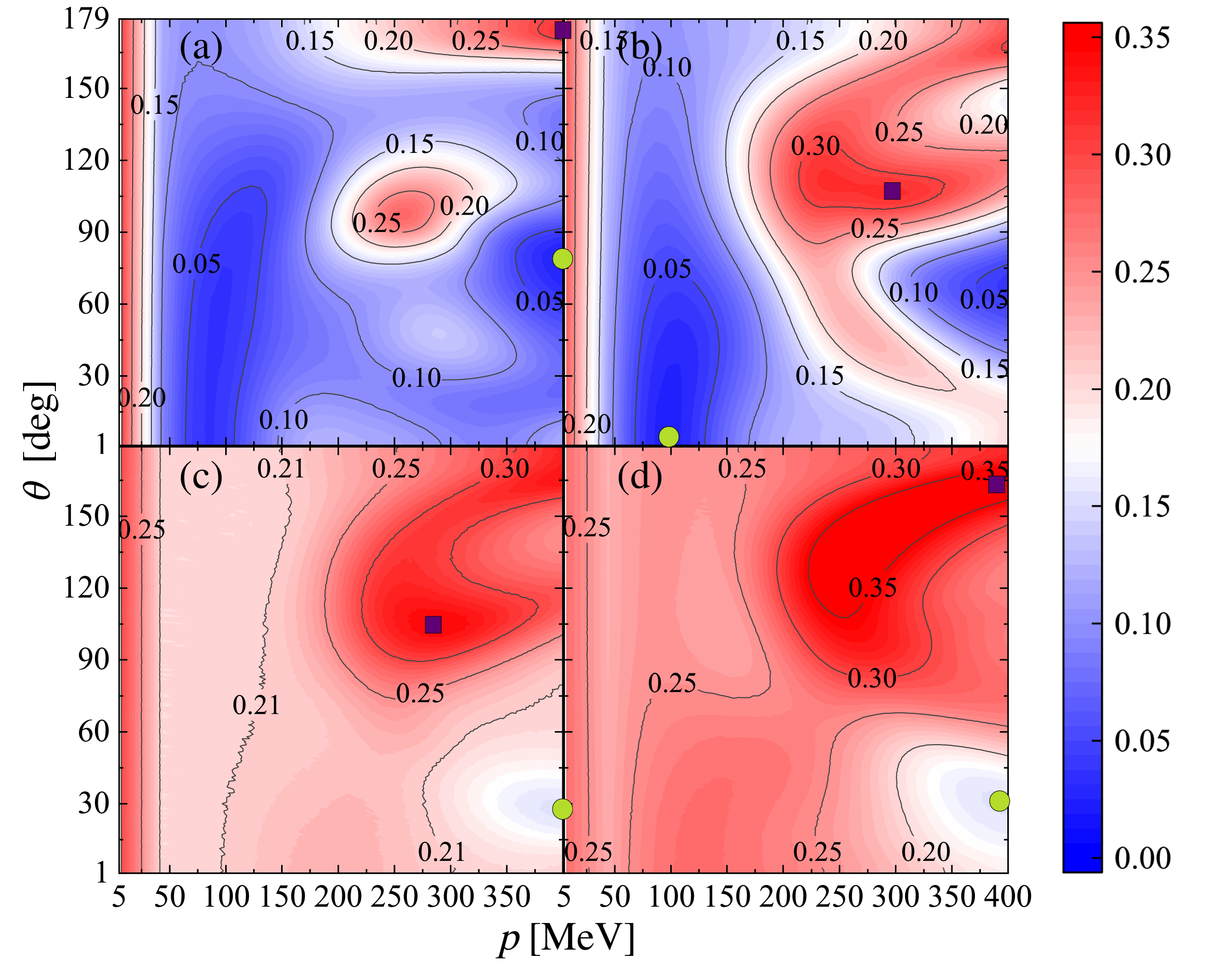}
		\caption{Momentum- and angle-dependent ensemble-averaged geometric quantum discord $\overline{\mathcal D}_G(p,\theta)$. Panels (a)--(d) correspond to averages over the same four input-state classes as in Fig.~\ref{fig:QIP_Iout}.}
		\label{fig:QIP_Dg}
	\end{figure}
	
	The ensemble-averaged geometric quantum discord $\overline{\mathcal D}_G(p,\theta)$ is presented in Fig.~\ref{fig:QIP_Dg}. This quantity probes nonclassical correlations beyond a purely classical description. As in the mutual-information case, intrinsic coherence or entanglement in the input ensemble raises both the peak nonclassical response and the background level of nonclassical correlation.
	
	
	
	For the diagonal mixed-state class, both the maximum and the minimum of $\overline{\mathcal D}_G$ occur in the high-momentum sector at $(399.6~\mathrm{MeV},174.0^\circ)$ and $(399.6~\mathrm{MeV},78.0^\circ)$. Even a purely classical input ensemble in the chosen spin basis can therefore acquire a limited amount of nonclassical output correlation at suitable kinematic points. The pure-state classes display much broader regions of elevated discord. The Schmidt-like subclass gives the strongest and most extended high-momentum large-angle enhancement.
	
	Table I gives a global quantitative comparison. Several extrema occur at identical or nearby kinematic points. In the separable pure-state class, the maxima of $\overline{\mathcal I}$ and $\overline{S}_L$ are close to each other, appearing at $(237.5~{\rm MeV},110.0^\circ)$ and $(241.4~{\rm MeV},111.0^\circ)$, respectively, while the maximum of $\overline{D}_G$ occurs in the same intermediate-angle region at $(297.2~{\rm MeV},106^\circ)$. In the general pure-state class, the maxima of $\overline{\mathcal I}$ and $\overline{D}_G$ coincide at $(284.3~{\rm MeV},104.0^\circ)$, and their minima also coincide at $(399.6~{\rm MeV},27.0^\circ)$. In the Schmidt-like entangled subclass, the maxima of $\overline{\mathcal I}$, $\overline{\mathcal N}$, and $\overline{D}_G$ all occur at $(390.1~{\rm MeV},164.0^\circ)$, with the maximum of $\overline{S}_L$ located at a nearby high-momentum backward angle. Similar coincidences are also found in the diagonal mixed-state class, where the extrema of $\overline{\mathcal N}$ and $\overline{D}_G$ are located in the same high-momentum region. These coincidences suggest that the enhanced-correlation regions are not artifacts of a particular correlation measure. Rather, they reflect kinematic regions in which the spin-space $M$ matrix enhances several aspects of the outgoing two-spin correlation structure simultaneously.
	
	A similar situation was reported recently in unpolarized \(pp\) scattering by Shen et al. \cite{Shen2025BellTriplet}, where the product-state entanglement power and the concurrence obtained from an unpolarized initial density matrix display a common local maximum around $E_{\rm lab}=151~{\rm MeV}$ and $\theta=90^\circ$. In that case, the common enhancement was traced to an approximate Bell-state transition operator that produces a nearly pure Bell-triplet state. Although the present \(np\) calculation does not imply the same single Bell-transition mechanism, the repeated coincidence of extrema in Table  ~\ref{table_1} indicates that selected kinematic regions of the np $M$ matrix also act as common sources of total correlations, nonclassical correlations, and bipartite entanglement.
	
	\begin{table}[htbp]
		\centering
		\caption{Extrema of the ensemble-averaged quantum-correlation measures for the four input-state classes.}
		\label{table_1}
		\begin{tabular}{l c c c c c}
			\hline\hline
			\textbf{} & {Observables} & {(a) Classical} &{(b) Separable} & {(c) General} & {(d) Schmidt-like} \\
			\hline
			\multirow{8}{*}{max} 
			& \multirow{2}{*}{$\overline{\mathcal I}$} & 1.370 & {1.404} & 1.487 & 1.630 \\
			&   & (399.6 MeV, 167.0$^\circ$) & {(237.5 MeV, 110.0$^\circ$)}  & (284.3 MeV, 104.0$^\circ$) & (390.1 MeV, 164.0$^\circ$) \\
			\cline{2-6}
			& \multirow{2}{*}{$\overline{S}_L$} & 0.492 & 0.317 & 0.343 & 0.386 \\
			&   & (397.2 MeV, 164.0$^\circ$) & (241.4 MeV, 111.0$^\circ$) & (300.3 MeV, 105.0$^\circ$) & (397.2 MeV, 164.0$^\circ$) \\
			\cline{2-6}
			& \multirow{2}{*}{$\overline{\mathcal N}$} & 0.357 & 0.389 & 0.399 & 0.429 \\
			&   & (399.6 MeV, 174.0$^\circ$) & (399.6 MeV, 173.0$^\circ$) & (399.6 MeV, 167.0$^\circ$) & (390.1 MeV, 164.0$^\circ$) \\
			\cline{2-6}
			& \multirow{2}{*}{$\overline{\mathcal D}_G$} & 0.315 & {0.317} & 0.343 & 0.386 \\
			&   & (399.6 MeV, 174.0$^\circ$) & {(297.2 MeV, 106.0$^\circ$)} & (284.3 MeV, 104.0$^\circ$) & (390.1 MeV, 164.0$^\circ$) \\
			\midrule
			\multirow{8}{*}{min} 
			& \multirow{2}{*}{$\overline{\mathcal I}$} & 0.185 & {0.180} & 0.786 & 0.793 \\
			&   & (399.6 MeV, 77.0$^\circ$) & (97.1 MeV, {2.0$^\circ$}) & (399.6 MeV, 27.0$^\circ$) & (399.6 MeV, 32.0$^\circ$) \\
			\cline{2-6}
			& \multirow{2}{*}{$\overline{S}_L$} & 0.304 & 0.023 & 0.152 & 0.155 \\
			&   & (375.4 MeV, 33.0$^\circ$) & (97.1 MeV, 6.0$^\circ$) & (399.6 MeV, 28.0$^\circ$) & (387.7 MeV, 32.0$^\circ$) \\
			\cline{2-6}
			& \multirow{2}{*}{$\overline{\mathcal N}$} & 0.024 & 0.217 & 0.284 & 0.282 \\
			&   & (399.6 MeV, 80.0$^\circ$) & (75.4 MeV, 4.0$^\circ$) & (377.9 MeV, 37.0$^\circ$) & (357.4 MeV, 31.0$^\circ$) \\
			\cline{2-6}
			& \multirow{2}{*}{$\overline{\mathcal D}_G$} & 0.022 & 0.025 & 0.154 & 0.155 \\
			&   & (399.6 MeV, 78.0$^\circ$) & (97.1 MeV, {2.0$^\circ$}) & (399.6 MeV, 27.0$^\circ$) & (392.5 MeV, 31.0$^\circ$) \\
			\hline\hline
		\end{tabular}
	\end{table}
	
	For $\overline{\mathcal I}$, $\overline{\mathcal N}$, and $\overline{\mathcal D}_G$, the Schmidt-like entangled subclass yields the largest maxima. The general pure-state class and the Schmidt-like subclass also substantially lift the minima. Intrinsic coherence and intrinsic entanglement therefore reshape the overall correlation landscape rather than merely enhancing isolated regions. At the same time, the four measures respond differently to the kinematics. The maxima of $\overline{\mathcal I}$ and $\overline{\mathcal D}_G$ migrate with the input-state class. The maxima of $\overline{\mathcal N}$ remain concentrated in the high-momentum backward-scattering region. The linear entropy behaves differently again, with the diagonal mixed-state class showing the largest values due to persistent reduced-state mixedness.

	It is also useful to clarify the difference between the present work and Ref.~\cite{Bai2023PRCmixed}. 
	Although several quantum-information quantities considered there are formally similar to those used here, the underlying scattering map is different. 	Ref.~\cite{Bai2023PRCmixed} used the \(S\)-matrix description, where the scattering process is treated as a unitary transformation in the relevant spin space. 
	In the present work, we use the kinematically resolved spin-space amplitude \(M(p,\theta)\) and construct the normalized conditional output state $\rho_{\rm out}(p,\theta)$.

	This \(M\)-matrix map is not a global trace-preserving \(S\)-matrix evolution, but a conditional spin-state map selected at fixed outgoing momentum and scattering angle. The input states are also different. Ref.~\cite{Bai2023PRCmixed} mainly considered partially polarized product mixed states, while the present work compares diagonal mixed states, separable pure states, general two-qubit pure states, and a controlled Schmidt-like entangled subclass within the same \(M\)-matrix framework. 
	Therefore, even when the same quantum-information measures are used, the two calculations represent different physical averages. 	The former characterizes the quantum correlations generated by an \(S\)-matrix evolution for a restricted class of polarized inputs, whereas the present calculation quantifies how different input-state classes respond to a fixed-angle conditional \(M\)-matrix scattering amplitude.

	\section{Conclusion}	
	We have studied quantum-information aspects of \(np\)  scattering within the spin-space $M$-matrix framework. The calculation was performed for four representative input-state classes, namely diagonal mixed states, separable pure states, general two-qubit pure states, and a special Schmidt-like entangled subclass. For each class, ensemble-averaged output mutual information, reduced-state linear entropy, negativity, and geometric quantum discord were calculated in the relative-momentum-- scattering angle plane.
	
	The results show that the conditional outgoing spin state is governed jointly by scattering kinematics and by the structure of the incoming quantum ensemble. Input classes with stronger intrinsic coherence or entanglement generally yield larger maxima and higher minima in the mutual information, negativity, and geometric quantum discord. The Schmidt-like entangled subclass gives the strongest overall response among the four classes. The diagonal mixed-state class gives the weakest response for these three correlation measures.
	
	The different observables encode complementary aspects of the scattering response. The maxima of the mutual information and geometric quantum discord shift with the input-state class and, in several cases, occur at the same or nearby kinematic points. This indicates that the strongest total-correlation regions are accompanied by strong nonclassical correlations. By contrast, the maximum of the negativity remains localized in the high-momentum backward-scattering region for all four classes. This suggests that this region is most favourable for generating or maintaining bipartite output entanglement. The reduced-state linear entropy follows a different trend. For mixed-state inputs it mainly reflects persistent subsystem mixedness, while in the pure-state sector it tracks output-state mixing associated with enhanced quantum correlations.
	
	The present analysis emphasizes the ensemble dependence of output correlations. It shows that the enhanced-correlation pattern obtained from product-state averaging does not exhaust the correlation structure of the $M$-matrix map. When the incoming ensemble is changed from product pure states to general pure states or to a controlled entangled subclass, both the magnitude and the location of the averaged response can change. This provides a broader view of the quantum-information content of \(np\) 
	 scattering.
	
	Overall, the output spin-correlation pattern cannot be attributed solely to scattering kinematics or solely to input-state structure. It emerges from the interplay between the two. This perspective may be useful for future studies that connect nuclear scattering observables with quantum-correlation measures. It may also help clarify how two-body entangling mechanisms enter more complicated nuclear systems.
	
	\section*{Acknowledgments}
	This work was supported by the National Natural Science Foundation of China under Grant No. 12475149 and by the Guangdong Basic and Applied Basic Research Foundation under Grant No. 2024A1515010911. 
	
	\section*{Data Availability}
	The code and data that support the findings of this work are available from the authors upon reasonable request and will be deposited in Zenodo upon publication~\cite{xie2026zenodo}.

	\bibliographystyle{apsrev4-2}
	\bibliography{qinprefs}
	
\end{document}